\pdfoutput=1
\documentclass[%
 reprint,
superscriptaddress,
 amsmath,amssymb,longbibliography,
 aps,
 prb,
]{revtex4-2}%
\usepackage{hyperref}
\usepackage{multirow}
\usepackage{booktabs}
\usepackage[dvipsnames]{xcolor}
\usepackage{dsfont}
\usepackage{nicefrac}
\usepackage{bm}
\usepackage{graphicx}
\usepackage{amsmath}
\usepackage{color}

\usepackage{tikz}

\newcommand{\nav}{\langle n \rangle}

\newcommand{\spin}{\sigma}
\newcommand{\ospin}{\overline{\sigma}}

\newcommand{\rv}{{\bf r}}

\newcommand{\pv}{{\bf p}}
\newcommand{\kv}{{\bf k}}
\newcommand{\qv}{{\bf q}}

\usetikzlibrary{arrows,positioning} 
\tikzset{
    photon/.style={decorate, decoration={snake}, draw=black},
    electron/.style={draw=black, postaction={decorate},
        decoration={markings,mark=at position .55 with {\arrow[draw=black,thick]{>}}}},
    gluon/.style={decorate, decoration={snake},draw=black}, 
    >=stealth',
    punkt/.style={
           rectangle,
           rounded corners,
           draw=black, very thick,
           text width=6.5em,
           minimum height=2em,
           text centered},
    pil/.style={
           ->,
           thick,
           shorten <=2pt,
           shorten >=2pt,}
}

\begin{document}

\title{Revisiting superconductivity in the extended one-band Hubbard model: pairing
via spin and charge fluctuations}
\author{Mercè Roig}
\altaffiliation{These authors contributed equally to this work.}
\affiliation{Niels Bohr Institute, University of Copenhagen, DK-2200 Copenhagen, Denmark}

\author{Astrid T. R\o mer}
\altaffiliation{These authors contributed equally to this work.}
\affiliation{Niels Bohr Institute, University of Copenhagen, DK-2200 Copenhagen, Denmark}
\affiliation{Danish Fundamental Metrology, Kogle All\'e 5, 2970 H{\o}rsholm, Denmark}

\author{P. J. Hirschfeld}
\affiliation{Department of Physics, University of Florida, Gainesville, Florida 32611, USA}

\author{Brian M. Andersen}
\altaffiliation{Correspondence should be sent to bma@nbi.ku.dk}
\affiliation{Niels Bohr Institute, University of Copenhagen, DK-2200 Copenhagen, Denmark}

\date{\today}

\begin{abstract}
The leading superconducting instabilities of the two-dimensional extended repulsive one-band Hubbard model within spin-fluctuation pairing theory depend sensitively on electron density, band and interaction parameters. We map out the phase diagrams within a random phase approximation (RPA) spin- and charge-fluctuation approach, and find that while $B_{1g}$ ($d_{x^2-y^2}$) and $B_{2g}$ ($d_{xy}$) pairing dominates in the absence of repulsive longer-range Coulomb interactions $V_{\rm NN}$, the latter induces pairing in other symmetry channels, including e.g $A_{2g}$ ($g$-wave), nodal $A_{1g}$ (extended $s$-wave), or nodal $E_u$ ($p$-wave) spin-triplet superconductivity. At the lowest temperatures, transition boundaries in the phase diagrams between symmetry-distinct spin-singlet orders generate complex time-reversal symmetry broken superpositions. By contrast, we find that boundaries between singlet and triplet regions are characterized by first-order transitions. Finally, motivated by recent photoemission experiments, we have determined the influence of an additional explicitly attractive nearest-neighbor interaction, $V_{\rm NN}<0$, on the superconducting gap structure. Depending on the electronic filling, such an  attraction boosts $E_u$ ($p$-wave) spin-triplet or $B_{1g}$ ($d_{x^2-y^2}$) spin-singlet ordering.

\end{abstract}

\maketitle

\section{Introduction}

The discovery of high-temperature superconductivity in the cuprates inspired fundamental theoretical studies of the origin of superconductivity from repulsive electron-electron interactions. A common approach to investigating such unconventional superconductivity arising purely from repulsive interactions starts from the one-band two-dimensional (2D) Hubbard Hamiltonian defined on a square lattice. A large number of theoretical works have explored superconducting Cooper pairing within this model~\cite{KohnLuttinger,Scalapino86,Miyake86,Gros1987,Kotliar1988,Hlubina99,Chubukov92,Zanchi96,maier00,Halboth00,Lichtenstein_2000,Honerkamp01,Guinea04,Tremblay_2005,Kancharla08,Raghu2010,Gull13,Staar14,Corboz_2014,Romer2015,Chen15,Deng2015,Zheng16,Simkovic16,Jiang_2018,Jiang_2019,Romer_PRR_2020,Qin_PRX,Chung_2020,Gong_2021}. These studies include unbiased exact methods on small clusters, and a wide range of different approximate schemes to reach results for the thermodynamic limit. The latter include e.g. weak-coupling methods and pairing driven by spin fluctuations. Such approaches to the problem generally find that while the $B_{1g}$ ($d_{x^2-y^2}$) regime of superconductivity occupies a substantial region of the phase diagram close to half filling and moderate next-nearest neighbor hopping, a rich phase diagram is exhibited for other doping levels and coupling strengths~\cite{Hlubina99,Raghu2010,Deng2015,Romer2015,Kreisel2017}. Recently, it was found that the formalism of spin-fluctuation-mediated pairing within the random phase approximation (RPA) compares well to results obtained within the dynamical cluster approximation (DCA)~\cite{Romer_PRR_2020}. The latter approach includes dynamical  self-energy effects, and can access the interaction regime where the Coulomb repulsion becomes comparable to the bandwidth, a regime inaccessible within RPA due to the magnetic Stoner instability.

Longer-range Coulomb interactions were not taken into account in the pairing problem in the works referenced above. Several other studies, however, have explored the resilience of $B_{1g}$ ($d_{x^2-y^2}$) superconductivity to longer-range Coulomb repulsion from a wide range of different techniques~\cite{Scalapino_1987,Plekhanov03,onari04,Onari_JPSJ2005,Raghu12,Senechal2013,Huang2013,Plonka2015,Reymbaut2016,Wolf_2018,Hutchinson_2020,jiangmaier2018}.
For example, recent studies within DCA find that the $d_{x^2-y^2}$ solution is robust towards inclusion of the nearest-neighbor (NN) Coulomb repulsion, as long as it is smaller than 50 \% of the onsite Coulomb repulsion~\cite{jiangmaier2018}. Similar conclusions were found by Refs.~\cite{Plekhanov03,Senechal2013,Reymbaut2016}. The resilience of $d_{x^2-y^2}$ superconductivity is ascribed mainly to the retarded nature of the pairing. The extended Hubbard model has also been treated within the fluctuation exchange (FLEX) approximation~\cite{onari04}, where it was found that the $B_{1g}$ ($d_{x^2-y^2}$) regime close to half filling persists even in the regime of sizable NN repulsion. As hole doping is increased, a region of triplet superconducting order sets in, before the system is driven into a charge density wave (CDW) phase. At even larger hole doping values, a $d_{xy}$ superconducting solution is present at all strengths of NN repulsion $V_{\rm NN}$. Finally, the effect of longer-range Coulomb repulsion on the superconducting instabilities has also been addressed in the limit of weak coupling~\cite{Raghu12,Wolf_2018}. In Ref.~\onlinecite{Raghu12} for example, the superconducting instability was determined by the asymptotically exact weak-coupling approach, and it was found that the $A_{2g}$ ($g_{xy(x^2-y^2)}$) solution dominates a large region around half filling. This poses an apparent contradiction to the FLEX results of Ref.~\onlinecite{onari04} which did not find a $g$-wave solution in any parameter range. Below we address this issue within the RPA spin-fluctuation approach by computing how the presence of longer-range interactions alters the superconducting instability in different interaction regimes. 

Previous RPA spin-fluctuation approaches to pairing in the Hubbard model including only onsite repulsion $U$ have mapped out the leading superconducting order as a function of band structure and interaction strength~\cite{Romer2015}. The obtained phase diagram agrees well with DCA and diagrammatic Monte Carlo simulations~\cite{Romer_PRR_2020}. All symmetry-allowed pairing states can be stabilized when varying the electron density and the band structure (e.g. by the next-nearest-neighbor (NNN) hopping $t'$)~\cite{Romer2015,Kreisel2017}. This includes a multi-nodal spin-triplet order near the van Hove filling where the band structure undergoes a Lifshitz transition. Thus, the expected role of including longer-range Coulomb interactions is mainly to modify the phase boundaries between distinct gap solutions, rather than generating entirely new ones. In addition, the detailed gap structure also changes due to the growing importance of higher harmonics with increasing $V_{\rm NN}$. 

At the phase boundary regions one expects linear superpositions of gap solutions at lower $T$. Below, we study several boundary cases and determine the possible coexistence of distinct gap solutions and the possibility of time-reversal symmetry breaking (TRSB). We note that longer-range Coulomb repulsion may be expected to be particularly relevant in cases of nearly degenerate pairing states, simply because it then tips the balance and becomes the determining factor for the preferred symmetry of the ground state condensate. An example of this scenario has been recently presented for a model relevant for Sr$_2$RuO$_4$, where theory indeed predicts near-degeneracy between several symmetry-distinct superconducting instabilities~\cite{Romer_2019,Romer_2022}. Specifically, $B_{1g}$ ($d_{x^2-y^2}$) superconductivity gives way to $B_{2g}$ ($d_{xy}$) pairing as a result of NN Coulomb repulsion~\cite{Romer2021_SRO}. Another example is found in the kagome lattice near its upper van Hove filling, where band structure effects "destructively interfere" for the onsite Coulomb term, again rendering NN repulsion important in deciding the basic pairing symmetry of the preferred superconducting solution~\cite{Romer_kagome}. 

The role of NN interactions in the pairing problem of the Hubbard model has been recently brought into the spotlight by angular-resolved photoemission spectroscopy (ARPES) on the doped cuprate chain-material Ba$_{2-x}$Sr$_{x}$CuO$_{3+\delta}$~\cite{Chen_science2021}. From an analysis of the detailed dispersion of the measured spinon and holon branches, it was inferred that this material exhibits a significant NN Coulomb {\it attraction}, comparable to the in-plane NN electron hopping integral. Such attraction may be generated via coupling to a phonon mode~\cite{Chen_science2021,Wang_2022}. The structural and chemical similarities between the cuprate chain compound and the actual high-$T_c$ cuprates has motivated theoretical studies of the role of NN attraction on the superconducting ground state~\cite{Qu_2021,Huang_2021,Jiang_PRB_2022,Chen_2022,Peng_2022}. In general, it is found that NN attraction tends to enhance superconducting correlations.

Here, we perform a systematic study of the effect of NN and NNN Coulomb interactions on the leading superconducting instabilities of the one-band square-lattice Hubbard model. We focus on the role of the interaction parameters, and investigate also different band structures and a wide range of electron filling concentrations. Initially, we focus on the case where all bare Coulomb interactions are repulsive and pairing is generated purely by effective attraction from both spin and charge fluctuations. Next, motivated by the recent ARPES results providing evidence for explicitly attractive NN coupling, we determine the leading superconducting instabilities in the case where only onsite repulsion $U$ is included in the spin-fluctuation approach, while an additional NN attraction $V_{\rm NN}<0$ enters directly in the particle-particle channel. In this framework, in addition to its ubiquitous onsite repulsive core, $U$ gives rise to an effective pairing in higher-angular momentum channels which coexists with the bare attraction. Thus, this approach is different from the case where both $U$ and $V_{\rm NN}$ enter as attractive channels at the bare level~\cite{Micnas_RevModPhys,Nayak_2018,Hutchinson_2020}.

\section{Model and Method}
We consider the one-band Hubbard model defined on a 2D square lattice given by 
\begin{eqnarray}
\hat H = -\sum_{i,j,\sigma} t_{ij} c^\dagger_{i\sigma}c_{j\sigma} -\mu \sum_{i,\sigma} n_{i\sigma} + \hat H_{\rm int}.
\end{eqnarray}
We include NN $t$ and NNN $t'<0$ hopping integrals. The NN hopping $t=1$ sets the energy scale.
The operator $c^\dagger_{i\sigma}/c_{i\sigma} $ denotes creation/annihilation of an electron with spin $\sigma$ at site $i$ and $n_{i\sigma}=c^\dagger_{i\sigma}c_{i\sigma}$. 
The interaction part of the Hamiltonian includes onsite $U$, NN $V_{\rm NN}$, and NNN $V_{\rm NNN}$ Coulomb interactions
\begin{eqnarray}
\hat H_{\rm int} &=&\frac{1}{2}\sum_{i,\sigma}U n_{i\sigma}n_{i\overline \sigma}+
\frac{1}{2}\sum_{i,\delta,\sigma,\sigma'}V(\delta) n_{i\sigma}n_{i+\delta,\sigma'}.
\end{eqnarray}
The vectors ${\bf \delta}\in\{\pm \hat {\bf x},\pm \hat {\bf y}\}~(\{\pm (\hat {\bf x}\pm \hat {\bf y})\})$ denote NN (NNN) lattice vectors with lattice constant $a=1$, and $V(\delta)$ corresponds to either $V_{\rm NN}$ or $V_{\rm NNN}$ depending on ${\bf \delta}$. By Fourier transformation we arrive at the free energy dispersion $\xi_\kv=-2t(\cos(k_x)+\cos(k_y))-4t'\cos(k_x)\cos(k_y)-\mu$, while the ($\qv=\mathbf{0}$) interacting part of the Hamiltonian takes the form
\begin{eqnarray}
\hat H_{\rm int} &=&\frac{U}{2N}\sum_{\kv,\kv',\sigma}
c_{\kv\sigma}^\dagger c^\dagger_{-\kv\overline\sigma}
c_{-\kv'\overline\sigma} c_{\kv' \sigma} \\ &+&\frac{1}{2N}\sum_{\kv,\kv',\delta,\sigma,\sigma'} V(\delta)
e^{-i\delta(\kv-\kv')}c_{\kv \sigma }^\dagger  c_{-\kv\sigma'}^\dagger c_{-\kv'\sigma' } c_{\kv'\sigma }.\nonumber 
\end{eqnarray}
We write the interaction Hamiltonian in the Cooper channel in the general form
\begin{eqnarray}
\hat H_{\rm int}&=&\sum_{\kv,\kv',\{\sigma_i\}}[V(\kv,\kv')]^{\sigma_1~\sigma_2}_{\sigma_3~\sigma_4}c_{\kv \sigma_1}^\dagger c_{-\kv \sigma_3}^\dagger
c_{-\kv' \sigma_2} c_{\kv' \sigma_4},
\label{eq:Hgeneral}
\end{eqnarray}
where $V(\kv,\kv')=V_{0}(\kv,\kv')+V_{\rm eff}(\kv,\kv')$. The bare interaction elements $V_0(\kv,\kv')$ are given by
\begin{widetext}
\begin{eqnarray}
&&\Big[V_{0}(\kv,\kv')\Big]^{\sigma~\sigma'}_{\sigma'\sigma}= U\delta_{\spin\ospin'}+2V_{\rm NN}[\cos(k_x-k_x')+\cos(k_y-k_y')]+4V_{\rm NNN}[\cos(k_x-k_x')\cos(k_y-k_y')], \label{eq:V01band} \\
&&\Big[V_{0}(\kv,\kv')\Big]^{\sigma~\sigma}_{\sigma'\sigma'}= -U\delta_{\spin\ospin'}-2V_{\rm NN}[\cos(k_x+k_x')+\cos(k_y+k_y')]-4V_{\rm NNN}[\cos(k_x+k_x')\cos(k_y+k_y')], 
\end{eqnarray}
where the second line is symmetry-imposed by the interaction Hamiltonian Eq.~(\ref{eq:Hgeneral}). 
The effective interaction $V_{\rm eff}(\kv,\kv')$ due to higher-order processes is given by
\begin{eqnarray}
&&\Big[V_{\rm eff}(\kv,\kv')\Big]^{\sigma_1~\sigma_2}_{\sigma_3~\sigma_4}=\sum_{\delta,\delta'}e^{-i\delta\kv}e^{i\delta'\kv'}\Big[\Big[W(\kv+\kv',\delta)\Big][\chi_{\rm RPA}(\kv+\kv',\delta,\delta')]\Big[W(\kv+\kv',\delta')\Big]\Big]^{\sigma_1~\sigma_2}_{\sigma_3~\sigma_4} \nonumber \\
&&\hspace{3cm}-e^{-i\delta\kv}e^{-i\delta'\kv'}\Big[\Big[W(\kv-\kv',\delta)\Big][\chi_{\rm RPA}(\kv-\kv',\delta,\delta')]\Big[W(\kv-\kv',\delta')\Big]\Big]^{\sigma_1~\sigma_4}_{\sigma_3~\sigma_2},
\label{eq:Veff1band}
\end{eqnarray}
where the matrices for the bare Coulomb interaction entering Eq.~(\ref{eq:Veff1band}) are 
\begin{eqnarray}
&&\Big[W(\qv,\delta=0)\Big]^{\spin~\spin}_{\spin' \spin'}=-U\delta_{\spin\ospin'} -2V_{\rm NN}[\cos(q_x)+\cos(q_y)]-4V_{\rm NNN}\cos(q_x)\cos(q_y), \label{eq:Vbarebub} \\
&&\Big[W(\delta=0)\Big]^{\spin ~\ospin}_{\ospin ~\spin}=U,  \hspace{.3cm} 
\Big[W(\delta \neq 0)\Big]^{\spin ~\overline \spin}_{ \overline \spin ~\spin}=V(\delta),  \hspace{1cm} 
\Big[W(\delta \neq 0)\Big]^{\spin ~ \spin}_{ \spin ~\spin}=V(\delta), 
\label{eq:Wmatrices}
\end{eqnarray}
\end{widetext}
where e.g. $V(\hat x)=V(\hat y)=V_{\rm NN}$ and $V(\hat x+\hat y)=V(\hat x - \hat y)=V_{\rm NNN}$. We return to the details of the effective pairing vertex below.

In order to determine the leading superconducting instabilities arising from spin and charge fluctuations, we solve
the linearized BCS gap equation
\begin{equation}
  -\frac{1}{(2\pi)^2}\int_{\rm FS} d \kv_f^\prime \frac{1}{|v(\kv_f^\prime)|} \Gamma_{s/t}(\kv_f,\kv_f^\prime)\Delta(\kv_f^\prime)=\lambda \Delta(\kv_f),
  \label{eq:LGE}
\end{equation}
where 
\begin{eqnarray}
\Gamma_{s/t}(\kv,\kv^\prime)=&&
[V_0(\kv,\kv')+V_{\rm eff}(\kv,\kv')]^{\sigma ~\overline \sigma}_{\overline \sigma ~\sigma}\nonumber \\
&&\mp
[V_0(\kv,\kv')+V_{\rm eff}(\kv,\kv')]^{\sigma ~\sigma}_{\overline \sigma ~\overline \sigma}\,,
\end{eqnarray}
is the spin-projected pairing kernel in the singlet ($s$) and triplet ($t$) channel. The wave vectors  at  the Fermi surface (FS) are denoted $\kv_f$ and $v(\kv_f)$ is the Fermi velocity.
The largest eigenvalue $\lambda$ and its associated eigenvector (gap function) $\Delta(\kv_f)$ correspond to the leading superconducting instability at $T_c$, but  subleading superconducting solutions are also obtained by this procedure.

In selected cases, we have also solved the full BCS gap equation in order to determine the low-$T$ gap structure near accidental degeneracy lines of the phase diagrams. This allows us to determine the order of the phase transition between different solutions, and to probe for spontaneously broken time-reversal symmetry. In the self-consistent solution of the full gap equation the mean-field gaps are labelled by the spin indices
\begin{align}
    [\Delta_{\bf{k}}]^{\sigma_1}_{\sigma_3} = \sum_{\kv',\sigma_2, \sigma_4} [V(\kv,\kv')]^{\sigma_1 ~\sigma_2}_{\sigma_3~\sigma_4} \langle c_{-\kv' \sigma_2} c_{\kv'\sigma_4} \rangle,
\end{align}
where $V(\kv,\kv')=V_{0}(\kv,\kv')+V_{\rm eff}(\kv,\kv')$. To classify the symmetries of the gap, we introduce the basis functions in the point group of the square lattice ($D_{4h}$). Therefore, we can rewrite the previous equation in terms of the allowed solutions for the gap depending on the form of the basis functions $g_{\kv}^\Gamma$,
\begin{eqnarray}
    [\Delta_\kv]^{\sigma_1}_{\sigma_3} = \!\! \sum_{\Gamma \in \textrm{IR}} g_{\kv}^\Gamma [\Delta_{\Gamma}]^{\sigma_1}_{\sigma_3},
    \label{eq:deltak_IR_selfconsistent}
\end{eqnarray}
where $\Gamma$ corresponds to the irreducible representations (IRs) of the point group $D_{4h}$ and
\begin{eqnarray}
     [\Delta_{\Gamma}]^{\sigma_1}_{\sigma_3}
    = \!\!  \sum_{\kv',\sigma_2,\sigma_4} \!\! [\tilde{V}^{\delta}]^{\sigma_1 ~ \sigma_2}_{\sigma_3 ~ \sigma_4} \; g_{\kv'}^\Gamma \; \langle c_{-\kv'\sigma_2} c_{\kv'\sigma_4} \rangle.
\end{eqnarray}
In the equation above, $[\tilde{V}^{\delta}]^{\sigma_1 ~ \sigma_2}_{\sigma_3 ~ \sigma_4}$ is the Fourier transform of $V(\kv,\kv')$ in Eqs.~(\ref{eq:V01band}-\ref{eq:Veff1band}) for the neighbor $\delta$. 
As an example, if we consider only nearest-neighbor interactions the pairing is given by
\begin{equation}
    [V(\kv,\kv')]^{\sigma_1 ~\sigma_2}_{\sigma_3~\sigma_4} = [\tilde{V}^{\rm NN}]^{\sigma_1 ~\sigma_2}_{\sigma_3~\sigma_4} \left( \cos(k_x-k_x')+\cos(k_y-k_y')\right),
\end{equation}
and therefore in this case the gap in Eq.~(\ref{eq:deltak_IR_selfconsistent}) can be written in terms of the basis functions
\begin{eqnarray}
    g_{\kv}^{A_{1g}} =& \cos k_x + \cos k_y, \qquad g_{\kv}^{E^x_{u}} = \sqrt{2} \sin{k_x}, \nonumber\\
    g_{\kv}^{B_{1g}} =& \cos k_x - \cos k_y, \qquad g_{\kv}^{E^y_{u}} = \sqrt{2} \sin{k_y}.
\end{eqnarray}
Following this procedure, we include interactions with up to 24 neighbors, as well as the on-site term.

The expression for the effective pairing Eq.~(\ref{eq:Veff1band}), obtained by evaluation of all bubble and ladder diagrams, is a matrix equation. The interaction matrices as well as the susceptibility become $2^2\cdot 5 \times 2^2\cdot 5 $ ($2^2\cdot 9 \times 2^2\cdot 9 $) matrices for NN interactions (NN and NNN) interactions, due to the spin degrees of freedom and the relevant range of the bare Coulomb interactions.
We introduce the generalized bare susceptibility by
\begin{eqnarray}
&&\chi_0(i\omega_n,\qv,\delta,\delta')\!
=\!\frac{1}{N}\sum_{\kv} e^{i\kv(\delta-\delta')}
 \frac{f(\xi_{\kv})-f(\xi_{\kv-\qv})}{i\omega_n+\xi_{\kv-\qv}-\xi_{\kv}},\nonumber \\
\label{eq:BareSus}
 \end{eqnarray}
 which includes the lattice vectors $\delta,\delta'$. This construction is similar to previous treatments of the pairing problem in the presence of longer-range Coulomb interactions~\cite{esirgenprb97,esirgenprb98}.
  The spin structure is included by construction of a susceptibility matrix $ \Big[\chi_0(i\omega_n,\qv,\delta,\delta')\Big]^{\sigma~ \sigma'}_{\sigma'~ \sigma}=\chi_0(i\omega_n,\qv,\delta,\delta')$, including $\sigma=\sigma'$ and $\sigma=-\sigma'$. 
The RPA susceptibility entering Eq.~\eqref{eq:Veff1band} takes the form
\begin{equation}
    \chi_{\rm RPA}(\omega_n,\qv,\delta,\delta')=[1-\chi_0(\omega_n,\qv) W(\qv)]^{-1}\chi_0(\omega_n,\qv,\delta,\delta'),
\end{equation} 
where summation in spin and $\delta$ indices is implicit. We evaluate the RPA susceptibility at zero energy, $\omega_n=0$, and temperature $k_BT=0.015$. This constitutes the main ingredient in the effective (static) pairing interaction, as seen from Eq.~(\ref{eq:Veff1band}). 

\begin{figure}[tb]
\centering
\includegraphics[angle=0,width=0.8\linewidth]{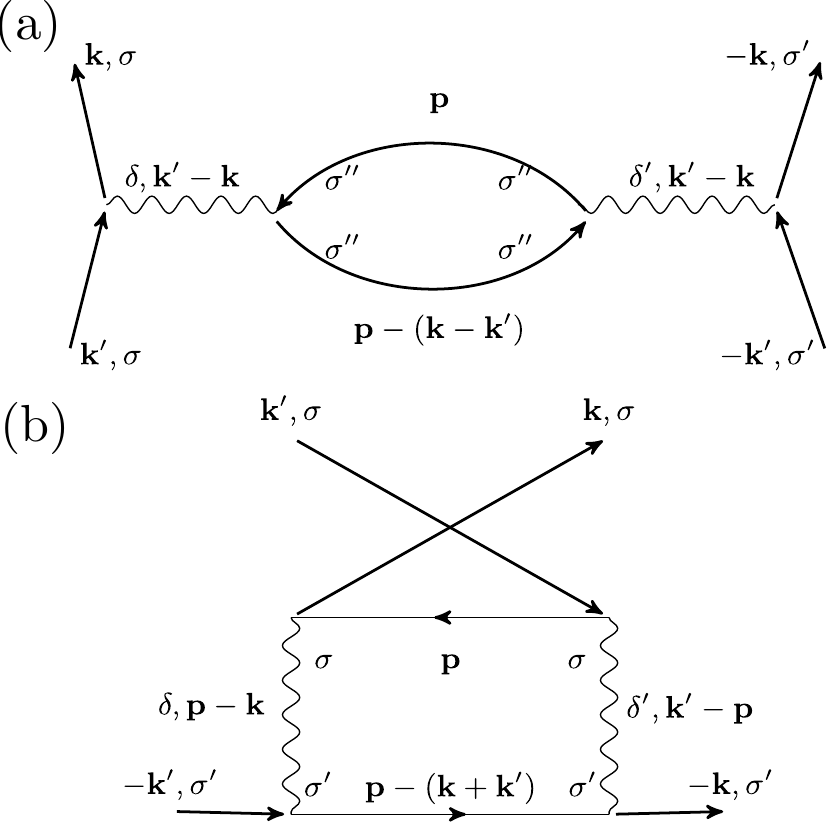}
\caption{Second order bubble (a) and ladder diagram (b). The real-space vector $\delta$ labels the range of the bare Coulomb interaction. In the bubble diagram (a), the $\delta$ summation can be performed independently of the $\pv$-sum in the bare susceptibility, while the ladder diagram (b) requires a summation in $\delta,\delta'$ which also involves the internal momentum label $\pv$.}
\label{fig:bubbleladder}
\end{figure}

The construction of the generalized susceptibility enables us to handle ladder diagrams in which some of the interaction vertices are longer-range Coulomb interactions. 
The bubble diagrams, on the other hand, are straightforward to sum up. Here, a direct summation in the real-space vector $\delta$ associated with each interaction line is possible because the momentum transfer is independent of the internal momentum label $\pv$ of the two propagators, as can be seen by inspection of the second-order diagram Fig.~\ref{fig:bubbleladder}(a). 
We note that, as a result of the sign convention in Eq.~(\ref{eq:Vbarebub}), the  bubble diagrams including only NN Coulomb repulsion $V_{\rm NN}$ give rise to an interaction proportional to the charge susceptibility: $ \chi_0(\qv)/[1+2V_{\rm NN}(\cos(q_x)+\cos(q_y))\chi_0(\qv)]$. At the same time, the spin structure of the matrix in Eq.~(\ref{eq:Vbarebub}) 
ensures that only an even number of bubbles is included in the case of onsite Coulomb interactions when the effective pairing for opposite spin electrons is considered.

Returning to the ladder diagrams, it can be seen from the second-order diagram of Fig.~\ref{fig:bubbleladder}(b)
that the interaction lines carry explicit reference to the internal momentum ${\pv}$ of the fermion propagators, as opposed to the bubble diagrams, where the tranferred momentum for each interaction line is simply given by $\kv'-\kv$.  This is why the additional structure  of the bare susceptibility including phase factors stated in Eq.~\eqref{eq:BareSus} is required  in order to sum up all the ladder diagrams. 

The physical spin and charge susceptibilities are obtained by
\begin{eqnarray}
\label{eq:spin_charge_sus}
 \Big[\chi_{\rm sp}^{\alpha \beta}(\qv,i\omega_n)\Big]&=& \frac{1}{N}\int_0^\beta d \tau e^{i \omega_n \tau}  \langle T_\tau S^\alpha (-\qv,\tau) S^{\beta} (\qv,0) \rangle \nonumber \\
= \frac{1}{4}&\sum_{\lbrace\sigma_i\rbrace}& \spin_{\sigma_1,\sigma_2}^\alpha \spin_{\sigma_3,\sigma_4}^{\beta} [\chi (\qv,\delta=0,\delta'=0)]^{\sigma_1; \sigma_2}_{\sigma_3;\sigma_4},\nonumber \\
\Big[\chi_{\rm ch}(\qv,i\omega_n)\Big]&=& \frac{1}{N}\int_0^\beta d \tau e^{i \omega_n \tau} \langle T_\tau n (-\qv,\tau) n (\qv,0) \rangle \nonumber \\
=\frac{1}{4} &\sum_{\sigma_1,\sigma_2}& [\chi (\qv,\delta=0,\delta'=0)]^{\sigma_1; \sigma_1}_{\sigma_2;\sigma_2},
 \end{eqnarray}
where the same result is obtained for the different spin channels $\alpha\beta \in\{xx,yy,zz\}$ since we consider neither magnetic order nor spin-orbit coupling~\cite{Romer2016,Romer_2019,merce2022}.

It is worth noticing that the interaction vertex also includes vertex corrections which are enabled by the presence of NN interactions. In the formalism where only the onsite interaction is non-zero such vertex corrections are absent due to the spin constraint. The presence of longer-range Coulomb interaction relieves this constraint and allows for additional diagrams of the form of vertex corrections, similar to the multi-orbital RPA formalism as discussed in e.g. Ref.~\onlinecite{altmeyer16}.

\begin{figure}[t]
 \centering
   	\includegraphics[angle=0,width=.95\linewidth]{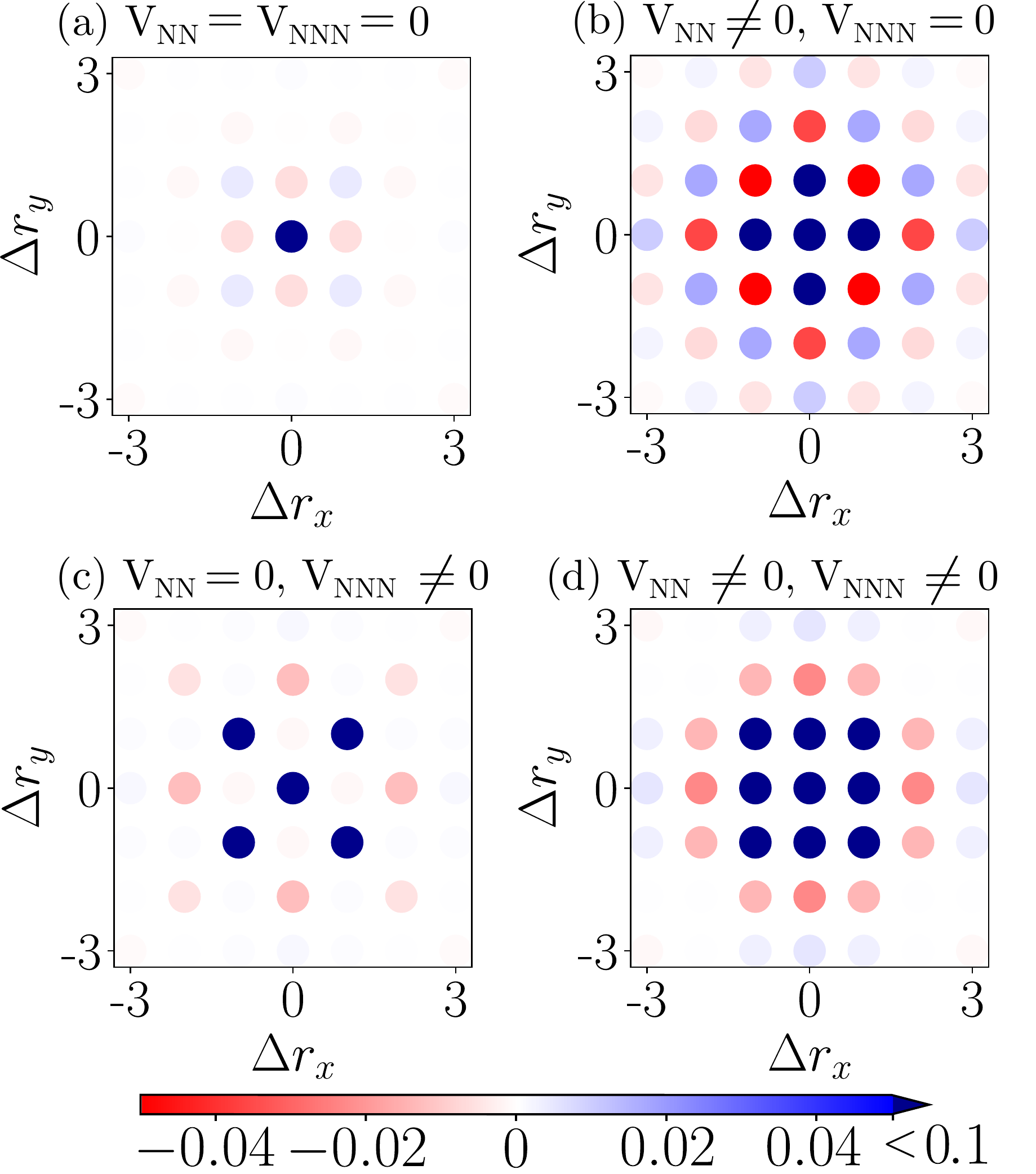}
   	\caption{Pairing interaction shown in real-space for opposite spin electrons in the singlet channel in the case of $\langle n \rangle=0.8$ and $t'=0$ obtained from Eqs.~(\ref{eq:V01band})-(\ref{eq:Veff1band}) with (a) only onsite $U$ repulsion ($U=0.5$), (b) same as (a) but including $V_{\rm NN}=0.25$, and (c) same as (a) but including $V_{\rm NNN}=0.125$. (d) The full pairing interaction vertex for the same parameters, i.e. $U=0.5$, $V_{\rm NN}=0.25$, and $V_{\rm NNN}=0.125$.}
   	\label{fig:RSpairing}
\end{figure}

Finally, for a classification of the different allowed gap solutions one needs to consider the irreducible representations (irreps) with two-dimensional basis functions of the relevant point group. In the present case, the point group of the square lattice is $D_{4h}$, which includes the irreps $A_{1g}$, $A_{2g}$, $B_{1g}$, and $B_{2g}$ in the singlet channel, and the two-dimensional irrep $E_u$ in the spin-triplet channel. Since we are considering a two-dimensional square lattice and a single band model, all solutions of the gap equation can be classified according to these irreps. Note, however, that a given solution may have contributions from several lattice harmonics, which is important for the total 
number of gap nodes. For example, the $B_{1g}$ irrep may correspond to a standard $d_{x^2-y^2}$ cuprate-like gap structure with 4 nodes on a $\Gamma$-centered Fermi surface, or it may involve contributions from an $i_{x^2(x^2-3y^2)^2-y^2(3x^2-y^2)^2}$-wave gap leading to 12 nodes on the Fermi surface. Both $d_{x^2-y^2}$ and $i_{x^2(x^2-3y^2)^2-y^2(3x^2-y^2)^2}$-wave orders transform according to the $B_{1g}$ irrep under the symmetry operations of $D_{4h}$~\cite{Wolf_2018,Romer2015}.

\begin{figure*}[t]
 \centering   	
 \includegraphics[angle=0,width=\linewidth]{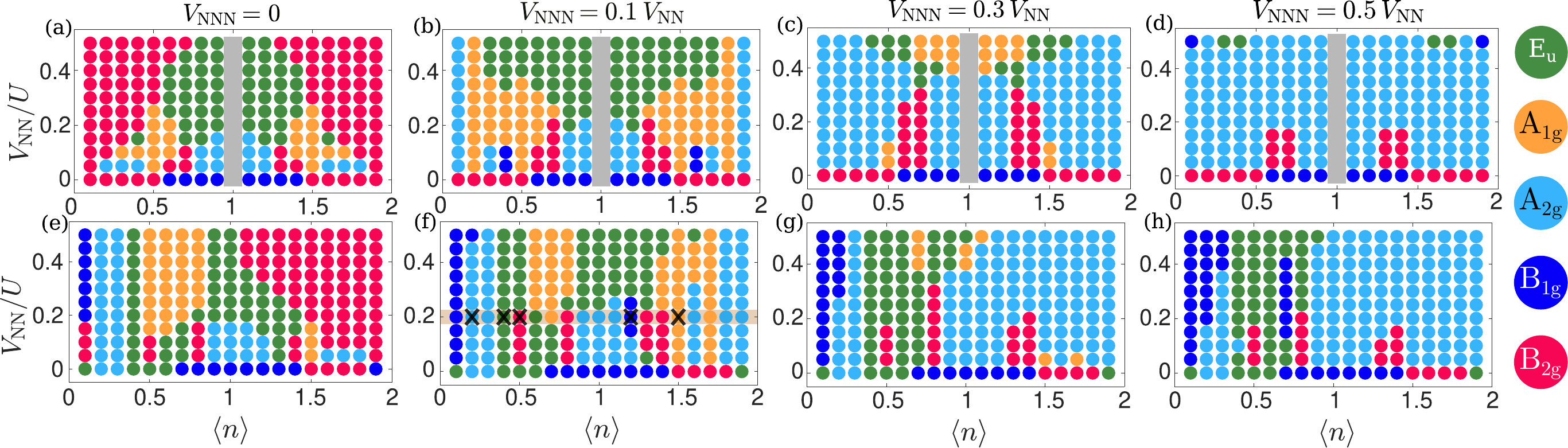}
\caption{Phase diagrams of the leading superconducting order as a function of the filling $\langle n \rangle$ and NN repulsion $V_{\rm NN}/U$, where $U=0.5$ is the onsite Coulomb repulsion. The color code indicates the irrep of the leading solution, and the center of each dot marks the actual parameter values of the computations. In panels (a)-(d) the NNN hopping is $t'=0$ and the system is thus symmetric with respect to electron and hole doping. Panels (e)-(h) are the same as (a)-(d) except for a finite $t'=-0.35$.
The NNN repulsion is zero in (a) and (e), while in (b,c,d)/(f,g,h) $V_{\rm NNN}=0.1\,V_{\rm NN},0.3\,V_{\rm NN},0.5\,V_{\rm NN}$, respectively. The gray color in the upper row indicates regions in which the spin or
charge susceptibility has diverged, causing a non-superconducting instability. The highlighted phase space points in panel (f) indicate the parameters used in Fig.~\ref{fig:gaps}.}
\label{fig:phasediagram}
\end{figure*}

\section{Results}
\subsection{Basic effect of longer-range Coulomb repulsion on pairing}
Before proceeding with a discussion of the obtained superconducting phase diagrams, it is instructive to consider the basic consequences of including longer-range repulsive Coulomb interactions in the pairing calculation. The long-range nature of the interaction is visualized as a function of inter-site spacing $\Delta {\bf r}=(\Delta r_x,\Delta r_y)$ in Fig.~\ref{fig:RSpairing}, providing an intuitive presentation of the extension of spin-fluctuation mediated pairing and the main effects of the longer-range Coulomb interaction. In the case of repulsive interactions, the longer-range terms act to "push out" the attractive sites leading to pairing in higher-angular momentum channels. To demonstrate this explicitly, we Fourier transform the pairing interaction $V(\kv,\kv')$, and analyse the pairing structure of the corresponding real-space Hamiltonian
\begin{equation}
\hat H_{\rm int}=\sum_{i,\Delta {\bf r},\sigma,\sigma'}
V(\Delta {\bf r})
c_{i\spin}^\dagger c_{i+\Delta {\bf r }\spin'}^\dagger
c_{i+\Delta {\bf r }\spin'} c_{i\spin}.
\end{equation} 
Figure~\ref{fig:RSpairing}(a) displays $V(\Delta {\bf r})$ in the case of onsite Coulomb repulsion only. Attraction occurs at NN sites which near half-filling produces a $B_{1g}$ ($d_{x^2-y^2}$) superconducting instability, a well-known result for the standard one-band Hubbard model~\cite{Scalapino86,Hlubina99,Romer2015,Romer_PRR_2020}. To visualize the effect of longer-range interactions we calculate the total pairing interaction of Eqs.~(\ref{eq:V01band})-(\ref{eq:Veff1band}) when including either NN or NNN repulsion. Repulsion between NN sites induces leading attractive couplings for NNN sites, and $\Delta {\bf r}=(\pm 2,0)$ and $(0,\pm 2)$, as shown in Fig.~\ref{fig:RSpairing}(b). NNN repulsion obviously disfavors attraction at NNN sites, and prefers attractive effective couplings at $\Delta {\bf r}=(\pm 2,0)$ and $(0,\pm 2)$, and $\Delta {\bf r}=(\pm 2,\pm 2)$, see Fig.~\ref{fig:RSpairing}(c). In Fig.~\ref{fig:RSpairing}(d) the full interaction as stated in Eq.~\eqref{eq:Veff1band}-\eqref{eq:Wmatrices} is shown. It includes all possible bubble and ladder diagrams, also mixed ones. As seen from Fig.~\ref{fig:RSpairing}(d), the bare interactions lead to an extended repulsive halo consisting of the closest eight neighbors. Thus, quantitatively even a modest longer-range Coulomb repulsion is enough to overwhelm the attraction generated by $U$. Overall attractive interactions generated by spin and charge fluctuations emerge beyond the closest eight neighbors. For the case shown in Fig.~\ref{fig:RSpairing}(d), the largest attractive sites are at $\Delta {\bf r}=(\pm 2,0)$, and $\Delta {\bf r}=(\pm 2,1)$ and symmetry-related sites. The latter tend to support an $A_{2g}$ ($g$-wave) solution whereas the former sites generate multi-nodal $B_{1g}$, $A_{1g}$ or $E_u$ spin-triplet solutions. As discussed in detail below, small differences in the pairing pattern become decisive for which superconducting instability is preferred at a given filling and band structure.

\subsection{Phase diagrams}
Figure~\ref{fig:phasediagram} displays an overview of the superconducting phase diagrams obtained within the current framework. More specifically, the phase diagrams shown in Fig.~\ref{fig:phasediagram} indicate the leading solution to the linearized gap equation Eq.~(\ref{eq:LGE}) for different electron filling $\langle n \rangle$, band structure and interaction parameters $V_{\rm NN}$ and $V_{\rm NNN}$. The panels \ref{fig:phasediagram}(a)-\ref{fig:phasediagram}(d) and \ref{fig:phasediagram}(e)-\ref{fig:phasediagram}(h) refer to cases with $t'=0$ and $t'=-0.35$, respectively. Figure~\ref{fig:phasediagram}(a) and \ref{fig:phasediagram}(e) display the superconducting phases in the absence of NNN Coulomb repulsion ($V_{\rm NNN}=0$), while Figs.~\ref{fig:phasediagram}(b)-\ref{fig:phasediagram}(d) and \ref{fig:phasediagram}(f)-\ref{fig:phasediagram}(h) show the results for increasing values of $V_{\rm NNN}$. Note that since the band is particle-hole symmetric at $t'=0$, the phase diagrams of Fig.~\ref{fig:phasediagram}(a)-(d) are symmetric with respect to electron and hole doping. For all results shown in Fig.~\ref{fig:phasediagram} we have fixed $U=0.5$. Generally, this value of $U$ is considerably below the critical interaction strength $U_c$ necessary for entering non-superconducting instabilities. The exception is the perfectly nested band with $t'=0$ close to half filling. Thus, $U_c$ varies considerably throughout the phase diagrams in Fig.~\ref{fig:phasediagram}, with e.g. $U_c=2.6$ at half-filling for the band with $t'=-0.35$. In the low-$U$ regime, the hierarchy of superconducting solutions for onsite interactions $U$ only has been shown to display overall agreement with DCA close to half filling, as discussed in Ref.~\onlinecite{Romer_PRR_2020}. In this regime different values of $U$ mainly shifts the overall amplitudes of the eigenvalues and while moderate changes in $U$ modify the phase boundaries slightly, it does not lead to significant changes to the phase diagrams of Fig.~\ref{fig:phasediagram} and is unimportant for the subsequent discussion.

Focusing first on the case with $t'=0$ displayed in Fig.~\ref{fig:phasediagram}(a)-\ref{fig:phasediagram}(d), we see that for onsite interactions only, the leading instability is either $B_{2g}$ ($d_{xy}$) at small to intermediate electron and hole fillings ($\nav \lesssim 0.5 $, $\nav \gtrsim 1.5 $) or $B_{1g}$ ($d_{x^2-y^2}$) for the other fillings ($0.5 \lesssim \nav \lesssim 1.5$), as seen from the $V_{\rm NN}=0$ line in all the panels Fig.~\ref{fig:phasediagram}(a-d). This is in agreement with previous results in the weak-coupling regime~\cite{Raghu12,Wolf_2018,Deng2015,Kreisel2017}. The momentum structure of these $B_{1g}$ and $B_{2g}$ solutions are well-described by the basic lowest harmonics, i.e. $\cos(k_x)-\cos(k_y)$ and $\sin(k_x)\sin(k_y)$, respectively, see also insets of Fig.~\ref{fig:Selfconsistent_cuts_currents}(a). Note that in the limit of very small $U$ (not shown), a spin-triplet solution may be present at the crossing point between the $B_{1g}$ and $B_{2g}$ solutions~\cite{Deng2015,Kreisel2017}. We have checked that indeed this triplet phase is there for very weak $U$ also within the present setup. 

\begin{figure*}[t]
 \centering
    	\includegraphics[angle=0,width=\linewidth]{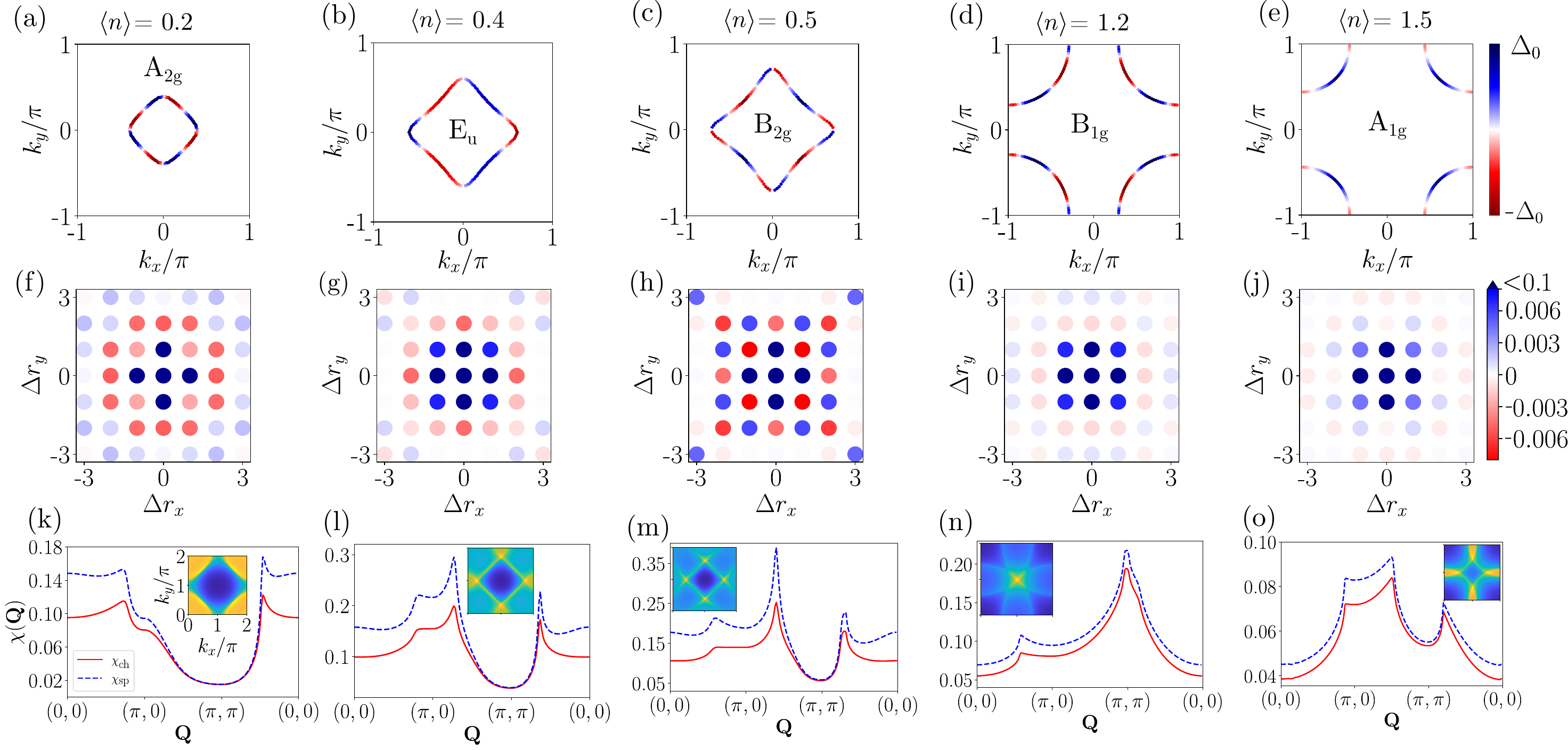}
\caption{(a)-(e) Momentum structure of the leading superconducting gap for electron fillings  $\nav=0.2,0.4,0.5,1.2,1.5$ in the case $U=0.5$, $V_{\rm NN}/U=0.2$, $V_{\rm NNN}/V_{\rm NN}=0.1$, with $t'=-0.35$, see also Fig.~\ref{fig:phasediagram}(f). Panels (f)-(j) display the real-space pairing interactions corresponding to the gap structures shown in (a)-(e). (k)-(o) Normal state RPA spin and charge susceptibilities shown by blue dashed lines and red full lines, respectively, calculated using Eq.~(\ref{eq:spin_charge_sus}) for the same parameters as the respective above panels. In the insets we show the full momentum dependence of the real part of the static spin susceptibility.}
\label{fig:gaps}
\end{figure*}

As longer-range repulsive interactions are included in the calculation of the effective pairing, very rich superconducting phase diagrams emerge as seen from the different panels in Fig.~\ref{fig:phasediagram}. In the vicinity of half-filling $\nav\simeq 1$, $B_{1g}$ ($d_{x^2-y^2}$) superconductivity is rapidly destroyed by NN repulsion since the scale of $V_{\rm NN}$ is large compared to the effective NN attraction induced only by $U$.
As seen from Fig.~\ref{fig:phasediagram}(a)-\ref{fig:phasediagram}(d), a transition to $A_{2g}$ ($g$-wave) order occurs irrespective of the size of $V_{\rm NNN}$. When $V_{\rm NNN}/U\lesssim 0.1$, a second transition from $A_{2g}$ ($g$-wave) to $E_u$ ($p$-wave) spin-triplet superconductivity is induced as the NN repulsion is increased beyond $V_{\rm NN}/U \simeq 0.15$. Figures~\ref{fig:phasediagram}(c)-\ref{fig:phasediagram}(d) and \ref{fig:phasediagram}(g)-\ref{fig:phasediagram}(h) reveal that the inclusion of sizable NNN repulsion promotes the $A_{2g}$ ($g$-wave) solution which dominates large regions of the phase diagram. However, a region of mainly $B_{2g}$ ($d_{xy}$) superconductivity persists for the case of $t'=0$ whereas the band structure relevant for $t'=-0.35$ prefers also the $E_u$ ($p$-wave) order in the hole-doped regime as seen from Fig.~\ref{fig:phasediagram}(g)-\ref{fig:phasediagram}(h). Overall the results shown in Fig.~\ref{fig:phasediagram} agree qualitatively with previous studies even though a direct comparison is not possible due to differences in the applied method or parameter values~\cite{Raghu12,Wolf_2018,Deng2015,Kreisel2017}.

\subsection{Momentum-dependent gap structures}

Next, we turn to a discussion of the detailed gap structures exhibited by the different solutions shown in Fig.~\ref{fig:phasediagram}. In general, as discussed in relation to Fig.~\ref{fig:RSpairing}, longer-range repulsion pushes attractive sites outward and creates an extended pairing structure in real-space, thereby equivalently generating significant contributions of higher harmonics within each irrep. Thus, longer-range Coulomb repulsion naturally generates additional nodes. In Fig.~\ref{fig:gaps} we display the different gap properties of typical solutions in Fig.~\ref{fig:phasediagram}. Specifically, the cases in Fig.~\ref{fig:gaps} correspond to $\nav=0.2,0.4,0.5,1.2,1.5$ with $U=0.5$, $V_{\rm NN}/U=0.2$, $V_{\rm NNN}/V_{\rm NN}=0.1$, and $t'=-0.35$. This parameter choice corresponds to a cut along the $V_{\rm NN}/U=0.2$ line in Fig.~\ref{fig:phasediagram}(f), see highlighted crosses. Figure~\ref{fig:gaps}(a)-\ref{fig:gaps}(e) display the gap structure on the Fermi surface, whereas Figs.~\ref{fig:gaps}(f)-\ref{fig:gaps}(j) show the corresponding real-space  pairing interactions. In panels \ref{fig:gaps}(k)-\ref{fig:gaps}(o) we show the related static spin and charge susceptibilities in momentum space. 

Starting from the low electron density region $\nav=0.2$, we see from panels \ref{fig:gaps}(f) and \ref{fig:gaps}(k) that the spin and charge susceptibilities, in conjunction with the repulsive bare interaction parameters, produce leading attractive sites with $\Delta {\bf r}=(\pm 2,1)$ and symmetry-related points. The lattice harmonic corresponding to these sites generates $g_{xy(x^2-y^2)}$-wave ($A_{2g}$) superconductivity, in agreement with the momentum structure of the gap shown in Fig.~\ref{fig:gaps}(a). At the larger electron concentration of $\nav=0.4$ the leading solution is a spin-triplet $E_u$ state. In this case, the Fermi surface and the interaction parameters conspire to produce the largest attractions on $\Delta {\bf r}=(\pm 2,0)$ and symmetry-related sites as seen from Fig.~\ref{fig:gaps}(g), essentially leading to a $\sin(2k_x)/\sin(2 k_y)$ $E_u$ gap structure in agreement with Fig.~\ref{fig:gaps}(b). From the susceptibilities in Fig.~\ref{fig:gaps}(l), one identifies a main nesting peak near $(\pi,0.2\pi)$ which connects same-sign regions in Fig.~\ref{fig:gaps}(b) due to the attractive sign of the pairing interaction in the spin-triplet sector. As discussed recently for UTe$_2$~\cite{Kreisel_triplet2022}, however, this imposes additional nodes and thus the gap structure is dominated by higher order harmonics. 
At $\nav=0.5$, as seen from Fig.~\ref{fig:gaps}(h), the main attractive sites are exhibited at the diagonal NNN positions $\Delta {\bf r}=(\pm 1,\pm 1)$. Thus, the modest bare NNN repulsion does not completely screen out attraction at these sites. 
This results in a leading spin-singlet $B_{2g}$ solution, see Fig.~\ref{fig:gaps}(c), where the additional longer-range attractions at $(\pm 2,\pm 2)$ seen from Fig.~\ref{fig:gaps}(h) generate a higher-order nodal structure. 
At $\nav=1.2$ where the Fermi surface resembles the cuprates, the main nesting is seen from Fig.~\ref{fig:gaps}(n) to be located close to the $(\pi,\pi)$ region, producing a $B_{1g}$ state. However, since the longer-range Coulomb repulsion has pushed the attractive sites beyond the NN sites, we find that the gap structure in  Fig.~\ref{fig:gaps}(d) displays large contributions for higher-order harmonics in the $B_{1g}$ channel. This produces additional nodes compared to the usual 4 symmetry-imposed nodes along the diagonal lines originating from the lowest harmonics. Finally, for an even larger filling factor of $\nav=1.5$, we show in Fig.~\ref{fig:gaps}(e) an example of an $A_{1g}$ gap solution with very extended Cooper pairing evident from the real-space pairing interaction shown in Fig.~\ref{fig:gaps}(j). We note that while we have discussed the obtained gap structures along a particular cut in Fig.~\ref{fig:phasediagram}(f), the gap structures shown in Fig.~\ref{fig:gaps}(a)-(e) are representative of their symmetry-equivalent partners in the other panels of Fig.~\ref{fig:phasediagram}, except for cases with vanishing $V_{\rm NN}$ and $V_{\rm NNN}$.

\subsection{Time-reversal symmetry breaking near degeneracy regions}

\begin{figure}[tb]
 \centering
\includegraphics[angle=0,width=\linewidth]{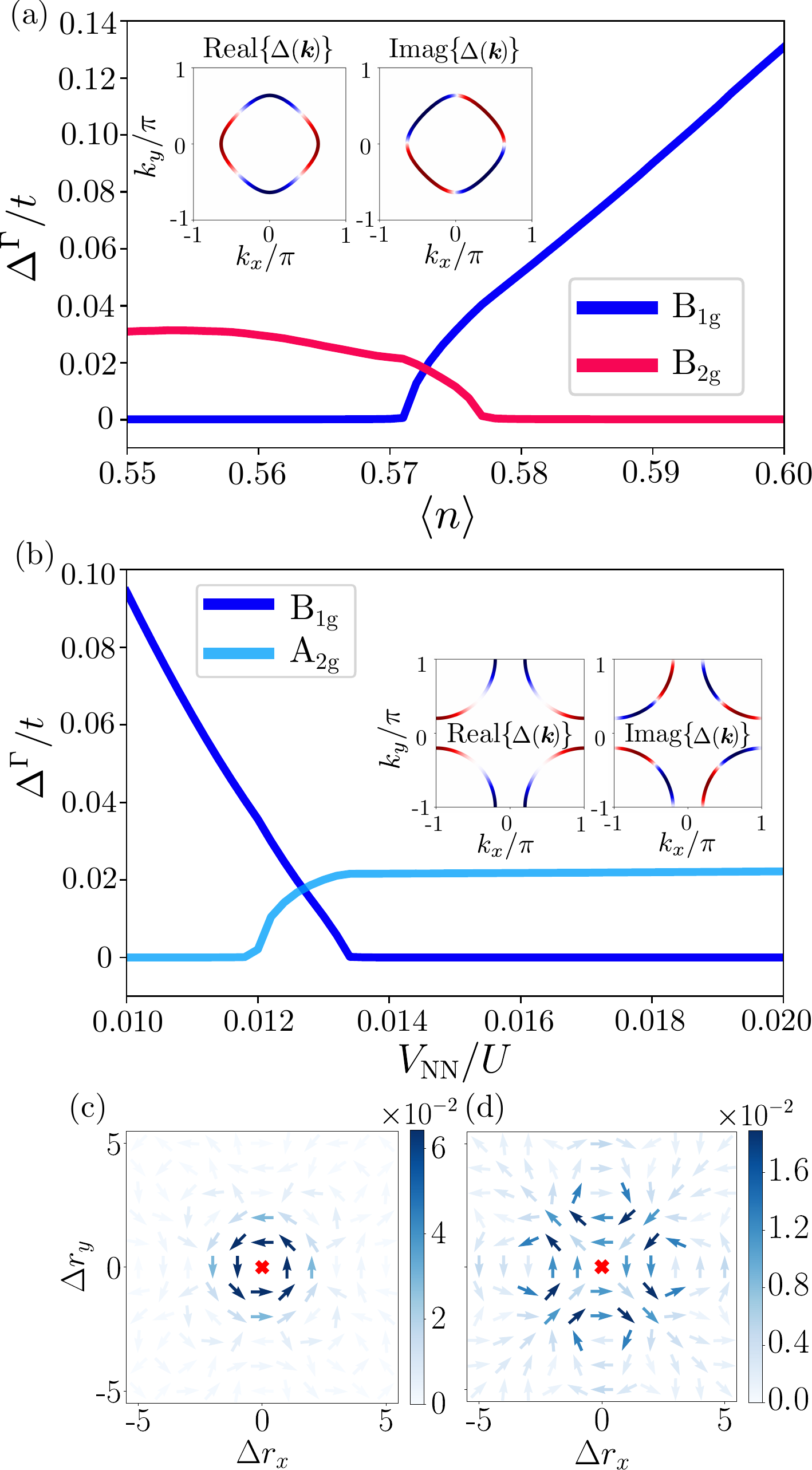}
\caption{Order parameters for (a) $\rm B_{\rm 1g}$ and $\rm B_{\rm 2g}$ and (b) $\rm B_{\rm 1g}$ and $\rm A_{\rm 2g}$ symmetry channels as a function of filling (a), or NN repulsion $V_{\rm NN}/U$ (b), with $U=0.5$. In (a) $V_{\rm NN} = V_{\rm NNN}= 0$ and $t'=0$, and in (b) $V_{\rm NNN}=0.5V_{\rm NN}$ at filling $\nav=1.0$, with $t'=-0.35$. The real and imaginary parts of the gaps are displayed in the insets at $\nav= 0.573$ (a) and at $V_{\rm NN}/U = 0.0126$ (b). Panels (c) and (d) display the current patterns induced around a nonmagnetic impurity placed at the center site (red cross) in the TRSB coexistence regions in (a) and (b), respectively. The currents are displayed in units of $et/\hbar a^2$.}
\label{fig:Selfconsistent_cuts_currents}
\end{figure}

From Fig.~\ref{fig:phasediagram} it is evident that the phase diagrams feature substantial areas of accidentally degenerate solutions to the linearized BCS gap equation. This poses the question: what is the nature of the superconducting condensates near these transition regions? To answer this question we have solved the full BCS gap equation along selected parameter cuts, crossing two symmetry-distinct solutions. All cases of transitions between spin-singlet and triplet regions are found to be first order. By contrast, Fig.~\ref{fig:Selfconsistent_cuts_currents} shows two examples where we zoom in on the transition between the two spin-singlet phases $B_{1g}$ and $B_{2g}$ ($B_{1g}$ and $A_{2g}$). The transitions correspond to increasing the filling for $V_{\rm NN}/U = 0$ in Fig.~\ref{fig:phasediagram}(a) (increasing $V_{\rm NN}/U$ for $\nav = 1$ in Fig.~\ref{fig:phasediagram}(h)). As seen both transitions are second order with a coexistence region in the crossover regime where the system stabilizes a complex TRSB phase, i.e. $B_{1g} + iB_{2g}$ and $B_{1g} + iA_{2g}$, respectively. The complex combination is verified explicitly from the solutions of the full BCS gap equation. In the insets of Fig.~\ref{fig:Selfconsistent_cuts_currents} we display the real and imaginary parts of the respective solutions.

TRSB in non-chiral superconductors can be exposed, for example, near inhomogeneities. Specifically, point-like disorder directly brings out the symmetry of the complex order parameter in localized supercurrents bound to the disorder sites~\cite{Lee2009,Maiti2015,Sigrist_2021,Clara2022}.  
To determine the impact of a single nonmagnetic impurity near the accidental degeneracy lines, we have solved the related real-space Bogoliubov-de Gennes equations and computed the resulting current densities between all NN and NNN bonds
\begin{align}
    \langle \Vec{J}^{~\rm NN}_{\rv} \rangle & = \sum_{\sigma} it \Big[ \hat{x} \langle c^\dagger_{\rv + \hat{x} \sigma} c_{\rv \sigma} \rangle + \hat{y} \langle c^\dagger_{\rv + \hat{y} \sigma} c_{\rv \sigma} \rangle - \rm H.c. \Big], \nonumber \\ 
    \langle \Vec{J}^{~\rm NNN}_{\rv} \rangle & = \sum_{\sigma} i \frac{t'}{\sqrt{2}} \Big[ (\hat{x}+\hat{y}) \langle c^\dagger_{\rv + (\hat{x} +\hat{y}) \sigma} c_{\rv \sigma} \rangle + \nonumber \\ 
    &\qquad \qquad ~ (\hat{x} - \hat{y}) \langle c^\dagger_{\rv + (\hat{x} - \hat{y}) \sigma} c_{\rv \sigma} \rangle - \rm H.c. \Big].
\end{align}
Figure~\ref{fig:Selfconsistent_cuts_currents}(c) and \ref{fig:Selfconsistent_cuts_currents}(d) show the resulting current patterns around a single 
impurity site (red cross). In agreement with earlier studies, loops of supercurrents are induced by the perturbation~\cite{Lee2009,Maiti2015,Sigrist_2021,Clara2022}. As seen, the induced pattern directly reflects the symmetries
of the condensate since Fig.~\ref{fig:Selfconsistent_cuts_currents}(c) (Fig.~\ref{fig:Selfconsistent_cuts_currents}(d)) corresponds to the cross product $B_{1g} \otimes B_{2g}=A_{2g}$ ($B_{1g}\otimes A_{2g}=B_{2g}$). Therefore, the current pattern exhibits the symmetries of the $ A_{2g}$ ($B_{2g}$) irreducible representation.
Thus, direct imaging of these current patterns allows for detailed exposure of the symmetries of the underlying condensate~\cite{Lin2016}. We stress that the emergent TRSB studied here is distinct from TRSB arising from impurity-induced spin-freezing caused by nonmagnetic disorder in correlated  superconductors \cite{Tsuchiura2001,ZWang2002,Zhu2002,Chen2004,Andersen2007,Harter2007,Andersen2007,Andersen2010,Schmid_2010,Gastiasoro2013,Gastiasoro2016,Martiny2019}. By including standard onsite Hubbard correlations in the particle-hole channel at the mean-field level, we have verified numerically that these two mechanisms of local TRSB  compete, i.e. local magnetic moments induced by $U$ suppress the impurity-induced local supercurrents.

\subsection{Consequences of additional nearest-neighbor attraction}

\begin{figure}[tb]
 \centering
\includegraphics[angle=0,width=\linewidth]{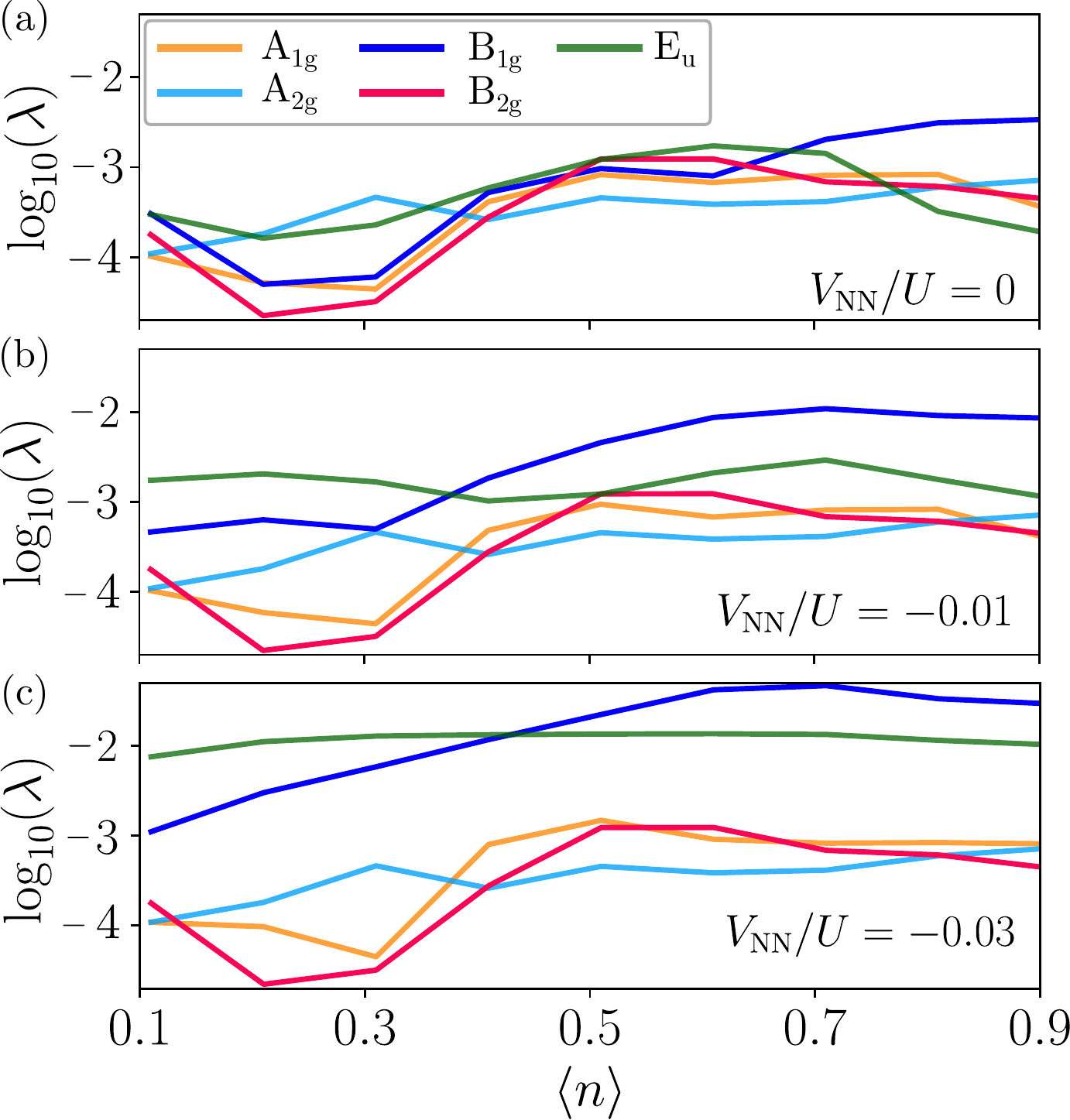}
\caption{Evolution of eigenvalues in distinct symmetry channels as a function of the electronic concentration $\nav$ in the case of $V_{\rm NN}/U = 0$ (a),  $V_{\rm NN}/U = -0.01$ (b) and $V_{\rm NN}/U = -0.03$ (c), where $U=0.5$ is the onsite Coulomb repulsion and $t'=-0.35$. Here, the attractive NN interaction is included at the bare level. For comparison, the effective NN attraction generated solely by $U$ (panel (a)) is of order $-0.01$.}
\label{fig:LGE_attractive_VNN}
\end{figure}

Finally, we turn to a discussion of the role of additional explicit (bare) attraction on the gap structures. This issue is motivated by the recent ARPES studies discussed in the Introduction, finding evidence for NN attraction possibly generated by appropriate phonon modes~\cite{Chen_science2021,Wang_2022}. We have mimicked this situation by calculation of the spin- and charge-fluctuation-generated pairing vertex from onsite $U$ repulsion only, and then added by hand an attractive NN coupling, i.e. imposed $V_{\rm NN}<0$ at the bare level of the pairing. In Fig.~\ref{fig:LGE_attractive_VNN} we show the leading eigenvalues of the linearized BCS gap equation as a function of electron concentration $\langle n \rangle$ and for increasing NN attraction. Figure~\ref{fig:LGE_attractive_VNN}(a) corresponds to a cut along the bottom edge ($V_{\rm NN}=0$) of Fig.~\ref{fig:phasediagram}(e). Figures~\ref{fig:LGE_attractive_VNN}(b) and \ref{fig:LGE_attractive_VNN}(c) display the same evolution of the eigenvalues as Fig.~\ref{fig:LGE_attractive_VNN}(a), but including negative $V_{\rm NN}$ in the pairing kernel. As seen, this simply raises the $E_u$ spin-triplet and $B_{1g}$ singlet irreps, which rapidly split off from the other solutions. Qualitatively, this is the expected behavior since attraction at the NN sites directly supports those irreps with a gap structure dominated by the corresponding lowest order harmonics. Quantitatively, however, Fig.~\ref{fig:LGE_attractive_VNN} reveals: 1) the crossover density of these two leading irreps, and 2) the significant eigenvalue enhancement already for relatively small explicit NN attraction. It is unclear to what extent these results directly relate to high-T$_c$ cuprate superconductivity, but the calculations do show that $B_{1g}$ ($d_{x^2-y^2}$-wave) superconductivity anchored by repulsive local Coulomb interactions may get its critical transition temperature significantly boosted by modest additional NN attraction.

\section{Discussion and conclusions}

We have determined the superconducting phase diagram of the repulsive one-band 2D Hubbard model within spin-fluctuation mediated superconductivity in the presence of longer-ranged Coulomb interactions. We have focused on how NN and NNN repulsive interactions alter the hierarchy of the leading  superconducting solutions. Our results align reasonably well with earlier works~\cite{Raghu12,Wolf_2018,Deng2015,Kreisel2017}, even though direct comparison is not possible due to the specific mechanism of spin and charge fluctuations mediating the pairing assumed here, incorporated via RPA summation of all bubble and ladder diagrams. The calculations reveal that longer-range interactions may strongly reshuffle the hierarchy of the leading pairing solutions. For example, as seen from the obtained phase diagrams (e.g. Fig.~\ref{fig:phasediagram}(e)), the amplitude of longer-range Coulomb repulsion alone may tune the ground state order from $B_{1g}\rightarrow A_{2g}\rightarrow E_u\rightarrow B_{2g}$.  In addition, the resulting gap structures tend to exhibit additional nodes due to the importance of higher-order lattice harmonics arising from the repulsive halo generated by the longer-range interactions. We have explored selected phase boundaries of the phase diagrams and determined the composite TRSB superconducting order near these regions. Transitions between symmetry-distinct spin-singlet order are found to be second order whereas spin-triplet/spin-singlet transitions are first order. We stress that these results are obtained in the static limit of the pairing kernel. Including retardation effects make short-range pairing more resilient to $V_{\rm NN}$ Coulomb repulsion~\cite{Plekhanov03,Senechal2013,Reymbaut2016,jiangmaier2018}. This may be important for  $d_{x^2-y^2}$ cuprate superconductivity where calculations estimate $V_{\rm NN}/U\sim 0.2$~\cite{Hirayama_2018}. Within the current framework, $B_{1g}$ ($d_{x^2-y^2}$) superconductivity is stable for similar $V_{\rm NN}/U$ ratios without retardation, but at interaction strengths $U$ close to the critical value $U_c$. (not shown explicitly here).

Finally, we have also determined the influence of an explicitly attractive NN interaction in the pairing kernel, in addition to the (onsite) Coulomb-interaction-generated effective vertex. Depending on the electronic concentration, the NN attraction boosts $E_u$ spin-triplet or $B_{1g}$ spin-singlet order, in both cases dominated by the lowest order harmonic as a result of the main attractive interaction residing on the NN sites. 
 
Our work highlights the richness of spin- and charge-fluctuation-mediated pairing in the 2D Hubbard model regarding the underlying band structure, but in particular as a function of changes in longer-range bare interaction parameters.  We expect that the results presented here will be of use to identify the possible states in new candidate unconventional superconductors with extended-Hubbard type correlations. It is interesting to note that the large number of phase boundaries present in the phase diagram provides many options for exotic TRSB condensates, where two symmetry-distinct orders of spin-singlet character combine into a complex order parameter. The symmetry of the combined order parameter is directly reflected in the locally induced currents around nonmagnetic disorder sites.

\section*{Acknowledgements} We acknowledge useful discussions with A. Kreisel and H. R\o ising. A.T.R. and B.M.A. acknowledge support from the Independent Research Fund Denmark grant number 8021-00047B. A.T.R. acknowledges support from the Danish Agency for Higher Education and Science.
M.R. acknowledges support from the Novo Nordisk Foundation grant NNF20OC0060019. P.J.H. acknowledges support from U.S. Dept. of Energy under DE-FG02-05ER46236.
 \bibliography{bib_1band}
\end{document}